# Unpacking Digital Sovereignty: A Three-Pillar Framework and a Research Agenda

*Angelos Kostis[1,2], Franz Strich[3], Bomi Kim[4], Pavlos Fafalios[1]*

*[1]Production Engineering & Management, Technical University of Crete*
*[2]Umeå School of Business, Economics and Statistics, Umeå University*
*[3]Business Administration and Economics, Paderborn University*
*[4]Stockholm School of Economics*

## Abstract

*Digital sovereignty has moved rapidly from policy jargon to global discourse, yet it lacks a solid conceptual foundation. Building on research from other disciplines, we propose a three-pillar framework (governance, infrastructure, and practice) to help IS scholars unpack how sovereignty is enacted, stabilized, or undermined across pillars. We argue that the most fruitful analytical work emerges at the intersections among the three pillars, where IS research is particularly well positioned to contribute. We introduce a local-distal axis that positions actions across scales of scope and proximity, and the notion of anchors, i.e., critical phenomena that cut across pillars and make the framework analytically tractable. We illustrate the framework using data as an anchor, showing how digital sovereignty depends on the alignment of governance, infrastructure, and everyday practice, and is undermined where they diverge. We close with a research agenda for IS scholars studying digital sovereignty empirically.*

## Introduction

This short paper aims to open a conversation by problematizing the notion of digital sovereignty. We propose a framework and outline avenues for future research. Digital sovereignty is defined here as "the practical capacity of an actor (an individual, community, organization, or nation) to make and enforce choices about how digital systems are sourced, procured, designed, governed, and used as well as to avoid, reshape, or exit dependencies that create risks to autonomy, rights, resilience, or collective values" (Nielsen et al., 2026). From a processual perspective, digital sovereignty encapsulates how entities are enacting their capacity to retain or regain control over their digital environments, and the evolution of those environments, to ensure competitiveness in the global market and social prosperity (Couture & Toupin, 2019; Pohle et al., 2025).

Few concepts have entered policy discussions and global discourse as rapidly as digital sovereignty (Burwell & Propp, 2026; Danet & Desforges, 2020; Floridi, 2020; Glasze et al., 2023; Pohle et al., 2025). Three reasons help explain this rapid uptake. First, digital infrastructures and data afford unprecedented power to the actors who control them (Hummel et al., 2021), which can potentially make dependence on foreign technology a strategic vulnerability and a constraint on freedoms (Braun & Hummel, 2025; Rainie et al., 2017). Second, policymakers have placed institutional emphasis on digital sovereignty as a precondition for growth, security, and democratic resilience. For instance, the European Commission has placed digital sovereignty at the center of its agenda, investing in sovereign cloud services, and in June 2026 it adopted a Communication on European Tech Sovereignty, accompanied by an EU Open Source Strategy (European Commission, 2026). A third reason is the pace at which digital technologies are built and deployed, which outpaces the institutions designed to regulate them, particularly when those technologies are deployed in regulatory and institutional landscapes different from those in which they were built. Together, these three intertwined developments have drawn growing attention to digital sovereignty from both researchers and practitioners.

Although digital sovereignty is becoming central to the agendas of policymakers and practitioners, there is no solid conceptual foundation on which to ground these efforts. Scholarship is only beginning to catch up, with calls for special issues in leading IS journals (Nielsen et al., 2026) and neighboring disciplines (Pohle et al., 2025), and with policy research that has started to specify what sovereignty requires of public administrations (Bryson et al., 2026). In this paper, we take a first step by offering three problematizations that can advance the nascent IS literature on digital sovereignty, drawing on the sociotechnical nature of the field and the data studies literature to argue how digital sovereignty can be studied. We provide an initial framework to guide future empirical investigations of digital sovereignty. We illustrate the framework through data, one of several possible anchors, showing how it can be operationalized and which assumptions of the current discourse need revisiting.

The nascent discourse on digital sovereignty requires further clarity on three fronts. First, the way digital sovereignty is theorized remains surprisingly narrow and fragmented. As reported in literature reviews on the topic, the meanings, components, and definitions of digital sovereignty differ dramatically with a predominant focus on state-level sovereignty and governance (Couture & Toupin, 2019; Pohle et al., 2025). For example, Hummel and colleagues (2021) document that the term carries a wide range of meanings across agents, contexts, and values, and authors frequently talk past one another by using the same word for different things, which causes conceptual ambiguity. Drawing on such critical voices, we argue that to unpack the nature of digital sovereignty and how it is built and enacted, we need to move beyond state-centric models that focus on state power to regulate and govern digital activities for national security, economic prosperity, and political agendas (see Couture et al., 2025; Pohle & Thiel, 2020; Santaniello, 2026). Consequently, our framework considers digital sovereignty as organizations', teams', and individuals' ability to control their own digital environments along well-defined dimensions that can guide a research program.

**Second**, understanding digital sovereignty requires looking not only at each pillar but at how the pillars interact. The three-pillar framework names the components, but components alone do not reveal the dynamics through which sovereignty is actually built or undermined. To trace these cross-cutting dynamics, we introduce the notion of anchors. Anchors can be defined as critical phenomena for an entity's digital environment that (i) cut across governance, infrastructure, and practice, and (ii) are at the core of how digital sovereignty is enacted or eroded. Here, we focus on data as one anchor and illustrate how the interacting dimensions of our framework help to unpack digital sovereignty. The nascent data sovereignty literature (Holmström & Magnusson, 2025; Moschko, Blazevic, & Piller, 2025; von Scherenberg, Hellmeier, & Otto, 2024) is helpful here as it concurrently drives attention to the role of data in achieving digital sovereignty. Whereas this body of research uses the term data sovereignty, its definition as the ability of entities to self-determine how data are produced, classified, stored, governed, shared, and used (e.g., Hummel et al., 2021; Jarke, Otto, & Ram, 2019) substantially overlaps with what this manuscript conceptualizes as digital sovereignty. Accordingly, data can be treated as an analytical anchor for examining digital sovereignty.

**Third**, the literature has placed disproportionate emphasis on isolation and strategic separation as the means by which sovereignty is achieved, an emphasis visible in policy-oriented and state-centric perspectives (Couture & Toupin, 2019). This framing emphasizes supreme authority, echoing colonial framings of sovereignty rather than autonomy (see Shrinkhal, 2021). We argue digital sovereignty requires a configuration in which differentiated actors develop along distinct trajectories while remaining close enough to interoperate, exchange knowledge, and learn from each other (see Holmström et al., 2024). We call this alternative configuration strategic embeddedness: the deliberate retention, diversification, or reconfiguration of dependencies, including the ability to exit them. The notion builds on research showing that global networks create asymmetries that central actors can exploit (Farrell & Newman, 2019), and it is consistent with recent policy analyses that describe digital sovereignty as the management of critical dependencies in ways that keep exit options, substitutions, and reconfigurations feasible (Bryson et al., 2026; Di Marco et al., 2025).

Drawing on Bratton (2016), who conceptualizes sovereignty as distributed across vertically differentiated layers of planetary-scale computation, we propose a three-pillar framework, i.e., (a) governance, (b) infrastructure, and (c) practice, as an analytical re-aggregation that captures how sovereignty is configured across, and enacted within, these layers. The framework maps the entanglements through which digital sovereignty is built and enacted (or not). Whereas each pillar can be explored independently, as has largely been the case so far, with the strongest focus on governance, the intersections among the three offer the richest ground for analysis, and this is where IS research can contribute most.

The contribution of this paper is threefold. First, building on research on digital sovereignty from other disciplines, it introduces a sociotechnical framework, illustrated with data, for analyzing how sovereignty is built, sustained, or eroded across the three pillars. Second, it provides a cross-sectional lens that integrates governance, infrastructure, and organizational practice as interdependent layers, specifying digital sovereignty as an emergent process of alignment rather than as a purely regulatory or technical outcome. Third, it articulates a research agenda for empirical IS work on digital sovereignty.

## A Three-Pillar Framework

To illustrate how the proposed framework can be put to use to unpack digital sovereignty, we proceed in three steps. We first describe the three pillars (i.e., governance, infrastructure, and practice) that together capture the layered structure through which digital sovereignty is built and enacted. We then introduce a local-distal axis that positions actions within each pillar along a continuum of scope and proximity, indicating whether interventions operate close to the immediate domain of control or at more peripheral points. Finally, we use data as an anchor to illustrate the framework in action and identify directions for future research.

### *Pillar I: Governance*

Governance encompasses the regulatory, legal, and policy mechanisms through which actors assert control over digital environments. It manifests across multiple levels: supranational and state instruments such as data protection law, AI regulation, and digital trade agreements, as well as organizational policies governing vendor selection, data handling, and AI use. It is the most visible pillar in the emerging public discourse, especially at the state level (Lambach & Oppermann, 2023). Governance defines the formal conditions under which digital systems and data flows are deemed legitimate and permissible, or strategically problematic. The EU AI Act, for instance, articulates ambitious sovereignty objectives, including human oversight of high-risk AI systems and obligations for providers of general-purpose AI models, that depend on infrastructural capacity and practice-level capabilities the regulation itself cannot create. Similar dynamics play out within organizations. A multinational firm's internal data flow policy that prescribes where customer data may be stored and processed across jurisdictions is a governance move that depends on infrastructure (whether in-region cloud capacity is actually available) and on practice (whether employees route data through the prescribed channels). While governance is only as effective as the infrastructures and practices that carry it, “the study of digital sovereignty as a set of infrastructures and socio-material practices has been comparatively neglected” (Musiani, 2022, p. 785).

### *Pillar II: Infrastructure*

Infrastructure is the material and ecosystemic substrate that enables digital sovereignty. We treat it as a layered stack with three analytically distinct sub-layers. First, *physical and energetic infrastructure* comprises the raw materials, energy capacity, hardware, and network elements on which all higher-order digital functions depend. Bria et al. (2025) document, for instance, that more than 80% of Europe’s digital technologies are imported, with critical chokepoints in raw materials, chips, and networks. Their EuroStack proposal calls for a layered technology stack spanning raw materials, chips, networks, IoT and devices, cloud, software, and data and AI as the material foundation for sovereignty. Beyond physical inputs, digital sovereignty also depends on an entity’s ability to adapt infrastructure under constraints such as energy availability, sustainability requirements, and network reach (Cruzes, 2026), especially given the high energy demands that AI workloads create (Bogmans et al., 2025). These pressures make clear that digital sovereignty depends on an entity’s capacity for stable and independent power generation. Because energy serves as the material precondition for any data-intensive operation, this layer requires more scholarly attention.

Second, *ecosystemic and entrepreneurial infrastructure* is the relational fabric of human capital, venture finance, research institutions, and standards bodies that allows sovereign digital capabilities to be produced and renewed (Bria et al., 2025). At this sub-layer, digital sovereignty requires the capability to build entrepreneurial ecosystems that can support the creation of solutions, services, and technologies needed for regaining control over an entity’s digital environment. Such configurations require alignment of scientific talent, public funding, regulatory market-shaping, and venture capital. The recent launch of Sferical AI in Sweden illustrates how an entrepreneurial ecosystem can be assembled to build critical sovereign infrastructure. Co-founded by AstraZeneca, Ericsson, Saab, SEB, and Wallenberg Investments, the initiative pools European capital, expertise, and cross-sector demand to build a sovereign AI compute platform that gives partner organizations access to frontier AI compute infrastructure independent of non-European cloud providers. The case also illustrates the logic of strategic embeddedness. The platform is built on NVIDIA hardware, showing that sovereignty at this sub-layer is rarely achieved through full self-sufficiency. Sferical AI differentiates Europe’s compute trajectory while retaining connections to global suppliers and partners.

Third, *data and semantic infrastructure* comprises the data architectures, shared standards, models, and ontologies through which data and its meaning travel across systems and organizations. A jurisdiction may have sovereign cloud capacity and a thriving entrepreneurial ecosystem, yet still be unable to make its data legible to itself or to its partners if the categories, classifications, and shared vocabularies that render data meaningful are designed elsewhere. Beyond syntactic interoperability, which enables the basic exchange of data, semantic interoperability, i.e., the capacity to exchange data with unambiguous, shared meaning, is a harder problem that depends on the development, alignment, and governance of shared ontologies (Auer, 2022; European Commission, 2020; Wilkinson et al., 2016). These ontologies are sociotechnical constructs: they embed assumptions, classifications, and relational structures, shaping how reality is represented and acted upon. Initiatives such as Gaia-X attempt to address this challenge by establishing federated data infrastructures with shared semantic standards for cross-organizational data exchange. Sovereignty at this sub-layer turns on the capacity to retain interpretive autonomy while participating in interconnected ecosystems (Bader et al., 2020).

### *Pillar III: Practice*

The third pillar, practice, captures the everyday enactment of digital sovereignty. It directs attention to the routines, habits, competencies, professional norms, and organizational dispositions through which claims to digital sovereignty are translated into concrete action or undermined in practice (Moschko et al., 2025). This perspective is necessary because digital sovereignty is not exhausted by state regulation or infrastructural capacity. It also depends on actors’ situated practices (Musiani, 2022), that is, on how they select tools, handle data, configure systems, follow security routines, assess vendors, procure technologies, and decide which platforms to rely on in daily work.

Practice shifts the analysis toward the situated level at which digital sovereignty becomes observable, both in individual choices about devices, software, privacy settings, and data disclosure, and in organizational routines around procurement, vendor selection, data governance, and security hygiene. This level matters because digital sovereignty claims increasingly extend beyond governments and states, while research on how nonstate actors engage with such claims in concrete situated practices remains limited (Pohle et al., 2025). Sovereignty in this sense concerns forms of autonomy and self-determination enacted by workers, managers, citizens, teams, organizations, and communities rather than asserted only through state authority (Couture & Toupin, 2019; Pohle & Thiel, 2020). Within organizations, much of this enactment falls to those who do data work day to day: data engineers, analysts, compliance officers, and frontline employees, whose decisions about classification, cleaning, sharing, and use are largely invisible to both governance instruments and infrastructure investment. Sovereignty is also negotiated in technical and engineering choices, where privacy and geopolitical concerns become aligned, or collide, in the design of specific systems (Fratini, 2025).

The practice layer also illustrates why formal compliance or sovereign infrastructure does not guarantee sovereign outcomes. An organization may be GDPR-compliant while its employees rely on personal devices, consumer cloud services, and non-European messaging tools for work. A government may invest heavily in sovereign cloud infrastructure while civil servants continue to default to US-hosted services out of habit, convenience, or lack of awareness. Generative AI services add a further layer, as platform owners can change the terms on which organizations build on and use them (Mayer et al., 2025). The picture becomes more tangled still as employees interact with partners across teams, organizations, and countries, each with its own history, norms, and standards. What looks coherent in a policy document scatters into countless small decisions that, though individually trivial, aggregate to constitute an organization's digital sovereignty (Jarke et al., 2019).

One emerging response to these challenges is the role of data stewards, who mediate between formal governance and situated practice by translating principles into workflows and cultivating norms within teams. Yet how such intermediary roles actually shape sovereignty in practice, and under what conditions they succeed or fail, remains an open empirical question.

Overall, governance defines formal authority, infrastructure provides the material and ecosystemic conditions for action, and practice determines whether sovereignty is enacted in situated use. The cross-layered dependency between governance, infrastructure, and practice illustrates why digital sovereignty cannot be advanced through any single pillar in isolation, and why strategic embeddedness operates across all three.

### *The Local-Distal Axis*

Each pillar operates simultaneously across multiple scales. To capture this, we propose positioning actions along a local-distal continuum that indicates their proximity to the focal objective and the extent to which they depend on external coordination and embeddedness. Governance, the most established articulation of digital sovereignty, illustrates the point: related activities can range from municipal data ordinances to international treaties. At the municipal level, cities may codify rights such as digital integrity or establish local data governance frameworks that regulate how data are collected and used within their jurisdiction. At the national level, governments implement regulatory regimes concerning data protection, cybersecurity, and platform accountability. At the supranational and international level, governance takes the form of cross-border agreements, standard-setting initiatives, and treaty-based coordination mechanisms that structure data flows and digital market access. The same continuum applies across the other two pillars: infrastructure ranges from local data centres to transnational supply chains, and practice ranges from individual tool choices to professional norms shaped by global communities. Digital sovereignty therefore emerges through continuous negotiation between local and distal forms, as actors rearticulate their objectives in response to changing dependencies. Movement along the local-distal axis is emergent, because data flows create dependencies that cannot be fully resolved through local control alone, nor fully delegated to distal arrangements.

## Putting the Framework into Motion

We use the notion of an anchor to make the framework actionable. As defined in the introduction, an anchor is a phenomenon that cuts across the three pillars and is central to an entity's digital environment. It provides a starting point from which actors can identify relevant dependencies, assess their exposure across the three pillars, and decide what actions are required. Focusing on an anchor enables organizations to begin with a specific concern and trace how it extends across adjacent domains.

The value of an anchor lies in its ability to hold together otherwise dispersed sovereignty-related actions. Once an organization selects an anchor, such as data, it can examine how this object is governed, which infrastructures it depends on, and which practices sustain or undermine it. Actions may remain close to the focal phenomenon or move towards more distal concerns. For example, an organization may begin with data and then address employee data-handling routines, standardization protocols for data flows, cloud-storage decisions, contractual arrangements with providers, cybersecurity requirements, or even upstream dependencies in hardware supply chains. These interventions differ in proximity and scope, yet they remain analytically connected because they are oriented towards the same anchor.

## Data as an Anchor

With the framework in place, we now demonstrate its analytical purpose by applying it to a concrete phenomenon. The framework is most useful when it illuminates phenomena that cut across all three pillars and along the local-distal axis.

**Data as a governance object.** Data is regulated by GDPR, sectoral legislation, the Data Governance Act, the Data Act, and a proliferating set of localization and cross-border transfer rules. IS scholarship has examined these questions primarily through the governance pillar by asking which rules exist, how they are enforced, and how organizations comply. Yet a governance-centered view remains incomplete. Data-related rules presuppose that organizations possess the infrastructural capacity to store, process, and transfer data in compliant ways, as well as the practical capabilities to classify, route, protect, and share data appropriately. However, even compliance-strong firms may display uneven sovereignty profiles and weak outward-facing capabilities.

**Data as an infrastructure object**. Data depends on physical servers, cloud architectures, hardware supply chains, network connectivity, energy provision, and compute capacity. Sovereign data infrastructure, including sovereign clouds, federated data spaces such as Gaia-X, and sovereign AI compute, is a condition for digital sovereignty. Beyond questions of where data is stored and through which hardware it flows, sovereignty also depends on whether actors can shape and govern the categories that render data meaningful. Standard data models, interoperability mechanisms, and shared ontologies are central here, because they shape whether data can move across actors while retaining clear and consistent meaning (Auer, 2022; Bader et al., 2020; European Commission, 2020; Wilkinson et al., 2016). Ontologies, classifications, and shared standards determine what data is before they determine what can be done with it. This raises questions the data sovereignty literature has only begun to engage: how ontology design processes unfold, who participates in setting semantic standards, how competing ontological frameworks interact, and how these dynamics shape inclusion, control, and dependency in digital environments (Lehuedé, 2022; Wiebe, 2026). Exit options also remain limited: a recent multi-layer assessment found that only 20% of 1,593 European software alternatives report an exit mechanism such as interoperability or data portability (Thabit Gonzalez et al., 2026).

**Data as a practice object**. Data is produced, classified, cleaned, and used through the everyday work of organizational actors, including data engineers, analysts, compliance officers, and frontline employees. Data quality, data ethics, and data sovereignty all depend on these mundane acts of data work, which are largely invisible to both governance instruments and infrastructure investment. Sovereignty in industrial firms, for example, is enacted through practices such as constituting the value of data, establishing evaluation mechanisms, developing standards and technological solutions, protecting data through anonymization, and building trust with external partners (Moschko et al., 2025). At the inter-organizational level, empirical work on data flows in digital innovation networks shows that such flows routinely stall at organizational seams (Holmström et al., 2024), undermining sovereignty even when governance and infrastructure are aligned.

The anchor also clarifies how sovereignty-related action moves along the local-distal continuum. Some data-related interventions remain close to the focal object, such as improving data classification, restricting access rights, strengthening privacy settings, or training employees in compliant data handling. Others move towards more distal dependencies, such as adopting sovereign cloud services, participating in federated data spaces, shaping sectoral data standards, securing AI compute, or reducing reliance on geopolitically exposed hardware and cloud providers. These actions differ in scope and proximity, yet they remain connected because they are organized around data as the focal sovereignty concern.

## A Research Agenda for Data Scholars

The framework we propose enables the generation of a focused research agenda for digital sovereignty as a multi-pillar phenomenon. Within each pillar, different clusters for potential future research are identified. The growing IS data-studies community (Alaimo & Kallinikos, 2021, 2022, 2024; Constantiou & Kallinikos, 2015; Jones, 2019), with its sociotechnical sensibility and engagement with data as both organizational resource and analytical object, is well placed to take this agenda forward.

### *Governance*

The governance pillar addresses how digital sovereignty is articulated through regulatory instruments and translated into organizational reality. A first cluster should focus on the translation problem (*Governance Cluster 1*; GC1), examining how mandates such as the GDPR, the Data Governance Act, the Data Act, the EU AI Act, and the proposed Cloud and AI Development Act are converted into internal policies, roles, and routines, and where this process breaks down. A second cluster could concern compliance and enforcement dynamics (GC2), analyzing the conditions under which organizations move from ceremonial compliance to substantive implementation. A third cluster may examine redistributive effects (GC3), focusing on how sovereignty obligations unevenly allocate costs and advantages across firms, sectors, and jurisdictions, thereby reshaping competitive dynamics. A fourth cluster could address cross-pillar interdependencies (GC4), investigating how governance outcomes depend on alignment with infrastructural capacities and organizational practices, and how misalignments undermine the realization of digital sovereignty.

### *Infrastructure*

The infrastructure pillar addresses the material and technical conditions under which digital sovereignty can be realized. A first cluster may focus on organizational adoption decisions (*Infrastructure Cluster 1;* IC1), examining how firms evaluate and select sovereign infrastructures such as sovereign clouds, federated data spaces, and sovereign AI compute under conditions of uncertainty regarding cost, performance, and geopolitical exposure. Ecosystem formation is a second cluster (IC2), which asks which configurations of public investment, regulatory market-shaping, and institutional support produce viable sovereign infrastructure and which lead to underutilized or stranded assets. A third cluster may examine coordination mechanisms (IC3), focusing on procurement, certification, and standards as infrastructural levers that connect ecosystem-level developments with organizational adoption. Finally, cross-pillar interdependencies (IC4) concern how infrastructural capabilities interact with governance requirements and organizational practices, and how misalignments limit the uptake and effectiveness of sovereign infrastructure.

### *Practice*

The practice pillar addresses how digital sovereignty is enacted in the situated activities through which data and related resources are produced, managed, and used. A first cluster may focus on the enactment of sovereignty in everyday data work (*Practice Cluster 1*; PC1). Digital sovereignty is realized through data work, the extensive and ongoing tasks of finding, curating, sharing, and using digital data, and through routine activities such as procurement, vendor selection, and data handling that can reinforce, undermine, or render sovereignty objectives invisible. Inter-organizational data flows form a second cluster (PC2), which examines how organizations manage fragmentation across silos with divergent standards and schemas, and how reducing epistemic uncertainty becomes a precondition for shared data resources and data commons required for AI systems. A third cluster could address coopetition and collective action problems (PC3), focusing on how tensions such as hidden agendas, knowledge misappropriation, free riding, and unequal value appropriation stall data sharing even where governance and infrastructure formally permit it. A fourth cluster could examine organizational competencies (PC4), analyzing how capabilities for exercising control over data assets are built, distributed, and contested across roles and functions. A fifth cluster could focus on values and relational resourcing (PC5), investigating how sovereignty practice is sustained through value-laden beliefs and the relational composition of resources under conditions of constraint and partial control.

## Conclusion

Digital sovereignty is among the most consequential strategic questions of the present moment, yet its theoretical treatment in IS remains thin. The three-pillar framework, the local-distal axis, and the illustrative example of using data as an anchor extend the conversation beyond governance to infrastructure and practice. Sovereignty, on this account, is a dynamic capacity, continuously produced through the looping interaction of the three pillars.